\documentclass[review]{elsarticle}
\usepackage{graphicx}  
\usepackage{dcolumn}   
\usepackage{bm}        
\usepackage{amssymb,amsmath}   
\usepackage{verbatim}
\usepackage[T1]{fontenc}
\usepackage{amsmath,amssymb,amsfonts}

 \usepackage{hyperref}
\hypersetup{colorlinks=true,linkcolor=magenta,anchorcolor=green,citecolor=cyan,filecolor=black,menucolor=black,urlcolor=brown}

\usepackage{cleveref}               

\usepackage{ifpdf}
\usepackage{color}

\usepackage{amsmath}
\usepackage{graphics}
\usepackage{mathtools}
\usepackage[usenames,dvipsnames]{xcolor}
\usepackage{epsfig}
\usepackage{epstopdf}
\usepackage{dcolumn}
\usepackage{tikz}
\usetikzlibrary{shapes.geometric, arrows}
\usepackage{upgreek}
\usepackage{setspace}
\usepackage{enumitem}
\usepackage{array,multirow,bigdelim}

\newcommand{\beq}{\begin{equation}}
\newcommand{\eeq}{\end{equation}}
\newcommand{\beqq}{\begin{equation*}}
\newcommand{\eeqq}{\end{equation*}}
\newcommand\beqa{\begin{eqnarray}}
\newcommand\eeqa{\end{eqnarray}}
\newcommand\beqaa{\begin{eqnarray*}}
\newcommand\eeqaa{\end{eqnarray*}}
\newcommand\bea{\begin{array}}
\newcommand\eea{\end{array}}

\usepackage{lineno,hyperref}
\biboptions{numbers,sort&compress}

\modulolinenumbers[5]

\journal{Physics Letter B}

\begin{document}

\begin{frontmatter}

\title{Post-Minkowskian Hamiltonians in modified theories of gravity\tnoteref{mytitlenote}}

\author{Andrea Cristofoli}
\address{Niels Bohr International Academy and Discovery Center,\\ Niels Bohr Insitute, University of Copenhagen,\\ Blegdamsvej 17, 2100 Copenhagen}

\ead{a.cristofoli@nbi.ku.dk}

\begin{abstract}
The aim of this note is to describe the computation of post-Minkowskian Hamiltonians in modified theories of gravity. Exploiting a recent relation between scattering amplitudes of massive scalars and potentials for relativistic point-particles we derive a contribution to post-Minkowskian Hamiltonians at second order in the Newton's constant coming from $\mathcal{R}^3$ modifications in General Relativity. Using this result we calculate the associated contribution to the scattering angle for binary black holes at second post-Minkowskian order, showing agreement in the non relativistic limit with previous results for the bending angle of a massless particle around a static massive source in $\mathcal{R}^3$ theories.
\end{abstract}


\end{frontmatter}

\newpage

\section*{Introduction}
\label{sec:headings}
The first detection of gravitational waves by the LIGO and Virgo collaboration, has opened up the possibility to test Einstein's theory of General Relativity at an unprecedented level, heralding a new era in fundamental physics \cite{Fundamental}. A central framework is the Effective One Body approach \cite{Buonanno, rewiev}, where information from Numerical Relativity and analytical approaches are combined in order to lead to improved gravitational wave templates. Among these several inputs, it has been recently suggested \cite{Damouruno, Damourdue} that also post-Minkowskian (PM) results, valid for weak gravitational fields and unbound velocities, can independently lead to improved modeling of bound binary dynamics. Given the growing results in post-Minkowksian physics \cite{Vines, Vines2, Guevara, Cheung, Chung, Bern, Antonelli, Cristofoli}, we would like to explore how contributions to post-Minkowskian Hamiltonians can be defined in modified theories of gravity. With no loss of generality, we here restrict ourselves on $\mathcal{R}^3$ modifications\footnote{These arise as further contributions to the Ricci scalar in the Einstein-Hilbert action, where the only non trivial modifications are given by $R^{\mu \nu}_{\: \: \alpha \beta}R^{\alpha \beta}_{\: \: \rho \sigma}R^{\rho \sigma}_{\: \: \mu \nu}$ and $R^{\mu \nu \alpha}_{\: \: \: \:\ \: \beta}R^{\beta \gamma}_{\: \: \: \nu \sigma}R^{\sigma}_{\: \: \mu \gamma \alpha}$.} to General Relativity \cite{Bueno1,Bueno2,Dunbar1,Dunbar2,stringa}. Recently, these have been studied in the context of scattering amplitudes \cite{Brand, Emond} leading to a post-Newtonian definition of the potential \cite{Iwa, Hol}. However, scattering amplitudes contain relativistic information that is lost in the passage to post-Newtonian point-particles potentials. We show how this can be restored defining a post-Minkwoskian potential in cubic theories of gravity, without restricting to the case of non relativistic point-particles. Using this result we derive the associated contribution to the fully relativistic scattering angle for binary black holes at second order in the Newton's constant. By then taking the non relativistic limit of one particle and the massless of the other, we are able to reproduce the bending angle recently calculated in \cite{Brand} for a massless particle around a static massive source.
\section{Higher derivative corrections in General Relativity}
A non-trivial modification of the one-loop scattering of massive scalars in cubic theories of gravity has been recently studied with amplitudes techniques in \cite{Brand, Emond}. In what follows we focus on the contribution given by $I_1 \equiv R^{\mu \nu}_{\: \: \alpha \beta}R^{\alpha \beta}_{\: \: \rho \sigma}R^{\rho \sigma}_{\: \: \mu \nu}$. As can be seen from \cite{Benincasa}, this arises as a non trivial modification to the usual Einstein-Hilbert action which for simplicity of discussion we will parametrize by an unknown coefficient $\alpha$ with the dimension of length squared, following \cite{Brand}. The associated classical information in the scattering of two massive scalars of masses $m_{1}$, $m_2$ has been calculated here \cite{Brand, Emond}. This is given by
\begin{figure}[h!]
\centering
\begin{equation}
\mathcal{M}^{\alpha}(p,q)=\begin{gathered}
 \includegraphics[width=2.8cm]{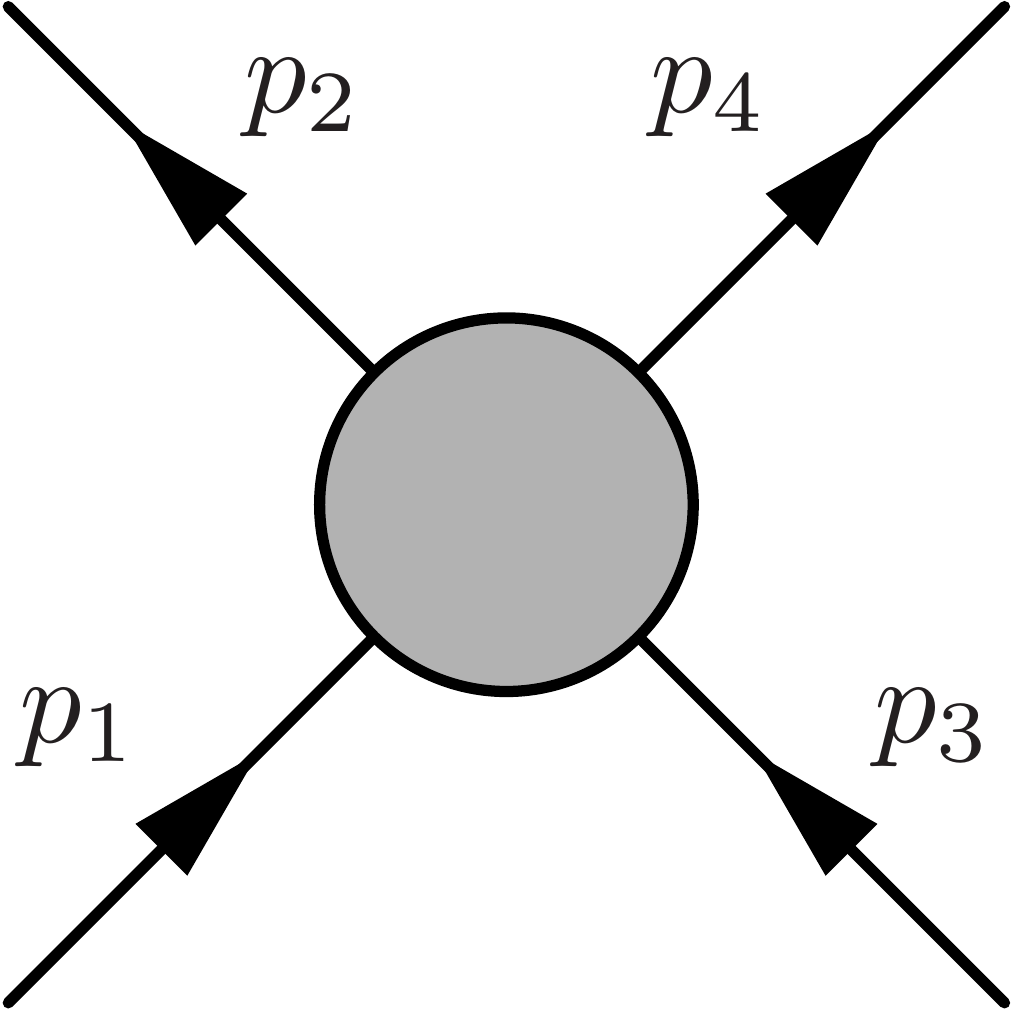}
 \end{gathered}
\end{equation}
\end{figure}
\begin{equation}
\label{Tra}
=\mathcal{D} \bigg[\mathcal{I}(m_1) \ c(m_1,m_2)+\mathcal{I}(m_2) \ c(m_2,m_1)\bigg]+...
\end{equation}
where, using $s=(p_1+p_3)^2$ and $t=(p_1-p_2)^2$, we have defined
\begin{equation}
\mathcal{D}= \frac{i\pi^2 G_N^2 \alpha^{2}}{\sqrt{E_1 E_2 E_3 E_4}}
\end{equation}
\begin{equation}\label{integra}
\mathcal{I}(m_j)=\int \frac{d^4k}{(2 \pi)^4}\frac{1}{(p_1-k)^2(p_3-k)^2(k^2-m^2_j)}
\end{equation}
\begin{equation}
c(m_i,m_j)=\frac{4t^2m^4_i}{(4m^2_i-t)^2}\bigg[\sum_{k=1}^{3}\beta_k(m_i,m_j)t^{(k-1)} \bigg]
\end{equation}
\begin{equation}
\beta_1(m_i,m_j)=2m^2_i \bigg[(m^2_i+m^2_j-s)^2-4m^2_im^2_j\bigg]
\end{equation}
\begin{equation}
\beta_2(m_i,m_j)=-3m^4_i + 2 m^2_i m^2_j+(m^2_j -s)^2
\end{equation}
\begin{equation}
\beta_3(m_i,m_j)=m^2_i - m^2_j +s
\end{equation}
We choose the center-of-mass frame and parametrize the momenta of the scattering particles as
\begin{equation}\begin{split} 
p^{\mu}_{1}&=(E_{1},\vec{p}\,)\,, \  \ \, p^{\mu}_{2}=(E_{1},\vec{p}\,')\, \\ 
p^{\mu}_{3}&=(E_{2},-\vec{p}\,)\,, \  p^{\mu}_{4}=(E_{2},-\vec{p}\,')\
\end{split}\end{equation}
\begin{equation} \label{para}
\vec{q}\equiv\vec{p}\,'-\vec{p}\,
\end{equation}
\begin{equation} 
|\vec{p}|=|\vec{p'}| \equiv p \quad , \quad  |\vec{q}| \equiv q
\end{equation}
We now proceed to define a post-Minkwoskian potential in the context of this modified theory of gravity using a recent relation between post-Minkowskian amplitudes and Hamiltonians \cite{Cristofoli}. The simplicity of this computation here lies in the lack of the Born subtraction, as there is no tree level amplitude to iterate that scales in the same way as (\ref{Tra}). We can thus define a post-Minkowskian potential to second order in $G_N$ and in the coupling $\alpha$ as
\begin{equation}\label{Fou}
V_{2PM}^{I_1}(p,r)=\int \frac{d^3q}{(2 \pi)^3} e^{i \vec{q} \cdot \vec{r}} \mathcal{M}^{\alpha}(p,q)
\end{equation}
By performing a proper $k^{0}$ integration on (\ref{integra}), the scalar triangle integral becomes \cite{Bohr, Cheung}
\begin{equation}
\mathcal{I}(m_j)=-\frac{i}{32 m_j q}+...
\end{equation}
where the ellipsis denote quantum contributions.\\
To leading order in $q$ the associated post-Minkowskian potential is\footnote{The reason we only keep the leading term in q is due by $\hbar$ counting. For a detailed analysis on how to restore the proper classical limit from an amplitude calculation see \cite{Oco}.}
\newline
\begin{equation}
V_{2PM}^{I_1}(p,r)=\frac{\pi^2 G^2_N \alpha^2}{32E_1 E_2}\int \frac{d^3q}{(2 \pi)^3} \bigg[\frac{c(m_1,m_2)}{m_1}+\frac{c(m_2,m_1)}{m_2} \bigg] \frac{e^{i \vec{q} \cdot \vec{r}}}{q}
\end{equation}
\begin{equation}
=\frac{\pi^2 G^2_N \alpha^2}{128E_1 E_2}\bigg(\frac{\beta_1 (m_1,m_2)}{m_1}+ \frac{\beta_1 (m_2,m_1)}{m_2}\bigg) \int \frac{d^3q}{(2\pi)^3}e^{i \vec{q} \cdot \vec{r}}q^3
\end{equation}
\begin{equation}
\label{pma}
V_{2PM}^{I_1}=\frac{3 \alpha^2}{32 E_1 E_2}\frac{G_N^2}{r^6}\bigg(\frac{\beta_1 (m_1,m_2)}{m_1}+ \frac{\beta_1 (m_2,m_1)}{m_2}\bigg)
\end{equation} 
\newline
In the non relativistic limit, our post-Minkwoskian potential reduces to
\begin{equation}
V_{2PM}^{I_1}(p,r)=\frac{3 \alpha^2}{4}\frac{G_N^2p^2}{r^6}\frac{(m_1+m_2)^3}{m_1m_2}+...
\end{equation}
in agreement with the post-Newtonian calculation in \cite{Brand}. 
For the sake of completeness we also report the post-Minkowskian contribution to the potential given by the remaining cubic term $R^{\mu \nu \alpha}_{\: \: \: \:\ \: \beta}R^{\beta \gamma}_{\: \: \: \nu \sigma}R^{\sigma}_{\: \: \mu \gamma \alpha}$. This has been recently calculated in \cite{Brand} as coming from the topological invariant $G_3=R^{\mu \nu}_{\: \: \alpha \beta}R^{\alpha \beta}_{\: \: \rho \sigma}R^{\rho \sigma}_{\: \: \mu \nu}-2R^{\mu \nu \alpha}_{\: \: \: \:\ \: \beta}R^{\beta \gamma}_{\: \: \: \nu \sigma}R^{\sigma}_{\: \: \mu \gamma \alpha}$. The result has been found equal to
\begin{equation}
\label{caro}
V_{2PM}^{G_3}(p,r)=\frac{12 \alpha^2G_N^2}{E_1E_2}\frac{m_1^2m_2^2(m_1+m_2)}{r^6}
\end{equation}
In a natural way, the same procedure for defining a post-Minkwoskian potential can be applied for more general modified theories of gravity.

\section{The scattering angle}
At second post-Minkowskian order in $G_N$, the Hamiltonian for a binary system of spinless binary black holes, including contributions from cubic gravity, is  given by
\begin{equation}
H_{2PM}^{\alpha}(p,r)=\sqrt{p^2+m_a^2}+\sqrt{p^2+m^2_b}+V_{2PM}(p,r)+V_{2PM}^{\alpha}(p,r)
\end{equation}
where $V_{2PM}(p,r)$ has been calculated here \cite{Cheung,Cristofoli}, being $V_{2PM}^{\alpha}$ the sum of (\ref{pma}) and (\ref{caro}).
Since the motion lies on a plane, we can introduce the following coordinates on the phase space $(r,\phi,p_r,p_{\phi})$ so as to express the momentum in the center of mass frame as
\begin{equation}
p^2=p^2_r+\frac{p^2_{\phi}}{r^2} \quad , \quad p_{\phi}=L
\end{equation}
being $L$ the angular momentum of the system, which is a conserved quantity.\\
The associated Hamilton-Jacobi equation is given by
\begin{equation}
\sqrt{p^2+m_a^2}+\sqrt{p^2+m^2_b}+V_{2PM}(p,r)+V_{2PM}^{\alpha}(p,r)=E
\end{equation}
with $E$ being the energy, another constant of motion.\\
By solving now in $p^2$ we can express the momentum in the center of mass frame as 
\begin{equation}
p^2=p^2(E,L,\alpha,r) \quad , \quad p^2=p^2_0+\frac{G_Nf_1}{r}+\frac{G^2_N f_2}{r^2}+\frac{G_N^2\alpha^2f_{\alpha}}{r^6}+...
\end{equation}
where the ellipsis denotes higher contributions in $G_N$ and
\newline
\begin{equation}
p^2_0=\frac{(p_1 \cdot p_2)^2-m^2_1m^2_2}{s} \quad , \quad f_1=-\frac{2c_1}{\sqrt{s}} \quad , \quad f_2=-\frac{1}{2\sqrt{s}}\bigg(\frac{c_{\bigtriangleup_a}}{m_a}+\frac{c_{\bigtriangleup_b}}{m_b} \bigg)
\end{equation}
\begin{equation}
f_{\alpha}=-\frac{3}{16E}\bigg(\frac{\beta_1 (m_1,m_2)}{m_1}+ \frac{\beta_1 (m_2,m_1)}{m_2}\bigg)-\frac{24m^2_1m^2_2(m_1+m_2)}{E}
\end{equation}
At this point, by considering the angular variable $\phi$, it is straightforward to derive the following expression for its total change during a scattering
\begin{equation}
\label{scat}
\Delta \phi = \pi + \chi \quad , \quad \frac{\chi(E,L)}{2}=-\int_{r_{min}}^{+\infty}dr \frac{\partial p_r}{\partial L}- \frac{\pi}{2}
\end{equation}
where $r_{min}$ is the positive root for $p_r=0$.\\
In order to evaluate (\ref{scat}) we proceeds perturbatively by expanding both the integrand and the extreme of integration in $G_N$, where
\begin{equation}
r_{min}=\frac{p_0}{L}+... \quad , \quad p_r=\sqrt{p^2_{0}-\frac{L^2}{r^2}}+...
\end{equation}
being the leading term of $r_{min}$ equivalent to the impact parameter $b$.\\
This expansion give rise to divergent integrals which can be handled only by means of the Hadamard Partie finie (Pf) of the latter as shown by Damour in \cite{Damourdue,Damourperi}. \\
Restricting to the contribution to (\ref{scat}) due to $\mathcal{R}^3$ one finds 
\begin{equation}
\frac{\chi_{2PM}^{\alpha}}{2}=-\frac{LG_N^2 \alpha^2f_{\alpha}}{2} \; \mathbf{Pf} \int_{r_{0}}^{+\infty}\frac{dr}{r^8}\bigg(p^2_0-\frac{L^2}{r^2} \bigg)^{-\frac{3}{2}}
\end{equation}
Changing variables to $u=\frac{1}{r}$ the integral becomes
\begin{equation}
\frac{\chi_{2PM}^{\alpha}}{2}=-\frac{G_N^2 \alpha^2f_{\alpha}}{2L^2} \; \mathbf{Pf} \int_{0}^{u_0} du \frac{u^6}{(u^2_0-u^2)^{\frac{3}{2}}} \quad , \quad u_0 \equiv \frac{1}{b} 
\end{equation}
The remaining integration is straightforward, leading to 
\begin{equation}
\frac{\chi_{2PM}^{\alpha}}{2}=\frac{15 \pi G_N^2 \alpha^2f_{\alpha}}{32L^2b^4} 
\end{equation}
\begin{equation}
\label{niuu}
\frac{\chi_{2PM}^{\alpha}}{2}=-\frac{45 \pi G_N^2 \alpha^2}{512 L^2 b^4 E}\bigg(\frac{\beta_1 (m_1,m_2)}{m_1}+ \frac{\beta_1 (m_2,m_1)}{m_2}+ 128 m^2_1m^2_2 (m_1+m_2) \bigg)
\end{equation}
\newline
Equation (\ref{niuu}) has to be considered as an additional contribution to the fully relativistic scattering angle at second order in $G_N$ coming from a cubic theory of gravity. In particular, by taking the non relativistic limit of our result with the additional condition $m_1=m$ and $m_2=0$, we have
\begin{equation}
\chi^{\alpha}_{2PM}=-\frac{45G^2_N \alpha^2 \pi m^2}{32b^6}+...
\end{equation}
which agrees with the non relativistic contribution derived in \cite{Brand} for the bending angle of a massless particle around a static massive source.\footnote{The authors in \cite{Brand} have used a convention for the deflection angle which differs by a minus sign compared to ours.}. In this case, the $G_3$ contribution to the potential is found to be absent for the bending angle of a massless particle, but not in the fully relativistic scattering angle of two massive particles as it can be seen from (\ref{niuu}).

\section*{Conclusion}
We have derived the post-Minkowskian contribution to relativistic point-particles Hamiltonians in modified theories of gravity. We have restricted ourselves to the case of $\mathcal{R}^3$ modifications, although similar changes are expected to appear also for $\mathcal{R}^2$ terms \cite{Vanhove1, Vanhove2,Alvarez}. The derived post-Minkowskian contribution, once expanded for small velocities, is in agreement with the recent post-Newtonian computation \cite{Brand}. The simplicity of the calculation has taken advantage of a recent relation between post-Minkowskian amplitudes and Hamiltonians for relativistic point-particles \cite{Cristofoli}.
Indeed, the computation has required no effective field theory matching as well as no need to known the operator reproducing the $\mathcal{R}^3$ modifications in an effective field theory of scalar fields. We have also derived an additional contribution to the fully relativistic scattering angle of black holes at second order in $G_N$ arising from $\mathcal{R}^3$, showing agreement in the non relativistic limit with a result derived  in \cite{Brand} for the bending angle of a massless particle around a static massive source. It would be interesting to systematically explore similar results in other alternative formulations of General Relativity. 
\newpage
\section*{Acknowledgements}
This work has been based partly on funding from the European Union's Horizon 2020 research and innovation programme under the Marie Sk\l{}odowska-Curie grant agreement No. 764850 (``SAGEX''). The author thanks Emil Bjerrum-Bohr, Poul Damgaard and Pierre Vanhove for useful conversations. The author is also grateful to Gabriele Travaglini, Andreas Brandhuber, Nathan Moynihan and Jan Plefka for comments and clarifications.


\section*{References}
\begin{thebibliography}{99}
\bibitem{Fundamental}
  B.~S.~Sathyaprakash {\it et al.},
  arXiv:1903.09221 [astro-ph.HE].
\bibitem{Buonanno}
  A.~Buonanno and T.~Damour,
  Phys.\ Rev.\ D {\bf 59} (1999) 084006
  doi:10.1103/PhysRevD.59.084006
  [gr-qc/9811091].
  
\bibitem{rewiev}
  T.~Damour,
  Int.\ J.\ Mod.\ Phys.\ A {\bf 23} (2008) 1130
  doi:10.1142/S0217751X08039992
  [arXiv:0802.4047 [gr-qc]].
  
\bibitem{Damouruno} 
  T.~Damour,
 ``Gravitational scattering, post-Minkowskian approximation and Effective One-Body theory,''
  Phys.\ Rev.\ D {\bf 94}, no. 10, 104015 (2016)
  doi:10.1103/PhysRevD.94.104015
  [arXiv:1609.00354 [gr-qc]].
  
 \bibitem{Damourdue} 
  T.~Damour,
 ``High-energy gravitational scattering and the general relativistic two-body problem,''
  Phys.\ Rev.\ D {\bf 97}, no. 4, 044038 (2018)
  doi:10.1103/PhysRevD.97.044038
  [arXiv:1710.10599 [gr-qc]].

\bibitem{Cheung} 
  C.~Cheung, I.~Z.~Rothstein and M.~P.~Solon,
  Phys.\ Rev.\ Lett.\  {\bf 121}, no. 25, 251101 (2018)
  doi:10.1103/PhysRevLett.121.251101
  [arXiv:1808.02489 [hep-th]].

\bibitem{Guevara}
  A.~Guevara, A.~Ochirov and J.~Vines,
  arXiv:1812.06895 [hep-th].
  
\bibitem{Vines}
  J.~Vines, J.~Steinhoff and A.~Buonanno,
  Phys.\ Rev.\ D {\bf 99} (2019) no.6,  064054
  doi:10.1103/PhysRevD.99.064054
  [arXiv:1812.00956 [gr-qc]].

\bibitem{Vines2}
  J.~Vines,
  Class.\ Quant.\ Grav.\  {\bf 35} (2018) no.8,  084002
  doi:10.1088/1361-6382/aaa3a8
  [arXiv:1709.06016 [gr-qc]].
\bibitem{Chung}
  M.~Z.~Chung, Y.~T.~Huang, J.~W.~Kim and S.~Lee,
  JHEP {\bf 1904} (2019) 156
  doi:10.1007/JHEP04(2019)156
  [arXiv:1812.08752 [hep-th]].
\bibitem{Bern} 
  Z.~Bern, C.~Cheung, R.~Roiban, C.~H.~Shen, M.~P.~Solon and M.~Zeng,
 ``Scattering Amplitudes and the Conservative Hamiltonian for Binary Systems at Third Post-Minkowskian Order,''
  arXiv:1901.04424 [hep-th].
 
\bibitem{Antonelli}
  A.~Antonelli, A.~Buonanno, J.~Steinhoff, M.~van de Meent and J.~Vines,
  ``Energetics of Two-Body Hamiltonians in Post-Minkowskian Gravity,''
  Phys.\ Rev.\ D {\bf 99} (2019) no.10,  104004
  doi:10.1103/PhysRevD.99.104004
  [arXiv:1901.07102 [gr-qc]].

\bibitem{Cristofoli}
  A.~Cristofoli, N.~E.~J.~Bjerrum-Bohr, P.~H.~Damgaard and P.~Vanhove,
  arXiv:1906.01579 [hep-th].


\bibitem{Dunbar1}
  D.~C.~Dunbar, J.~H.~Godwin, G.~R.~Jehu and W.~B.~Perkins,
  Phys.\ Lett.\ B {\bf 771} (2017) 230
  doi:10.1016/j.physletb.2017.05.052
  [arXiv:1702.08273 [hep-th]].
  
\bibitem{Dunbar2}
  D.~C.~Dunbar, J.~H.~Godwin, G.~R.~Jehu and W.~B.~Perkins,
  Phys.\ Lett.\ B {\bf 780} (2018) 41
  doi:10.1016/j.physletb.2018.02.046
  [arXiv:1711.05526 [hep-th]].
  
\bibitem{stringa}
  R.~R.~Metsaev and A.~A.~Tseytlin,
  Phys.\ Lett.\ B {\bf 185} (1987) 52.
  doi:10.1016/0370-2693(87)91527-9
  
\bibitem{Bueno1}
  P.~Bueno and P.~A.~Cano,
  Phys.\ Rev.\ D {\bf 94} (2016) no.10,  104005
  doi:10.1103/PhysRevD.94.104005
  [arXiv:1607.06463 [hep-th]].
  
\bibitem{Bueno2}
  P.~Bueno and P.~A.~Cano,
  Phys.\ Rev.\ D {\bf 94} (2016) no.12,  124051
  doi:10.1103/PhysRevD.94.124051
  [arXiv:1610.08019 [hep-th]].


\bibitem{Brand} 
  A.~Brandhuber and G.~Travaglini,
 ``On higher-derivative effects on the gravitational potential and particle bending,''
  arXiv:1905.05657 [hep-th].
  
\bibitem{Emond}
  W.~T.~Emond and N.~Moynihan,
  arXiv:1905.08213 [hep-th].
  
\bibitem{Iwa}
  Y.~Iwasaki,
  Prog.\ Theor.\ Phys.\  {\bf 46} (1971) 1587.
  doi:10.1143/PTP.46.1587
  
  
\bibitem{Hol}
  B.~R.~Holstein and A.~Ross,
  arXiv:0802.0716 [hep-ph].
\bibitem{Benincasa}
  P.~Benincasa and F.~Cachazo,
  arXiv:0705.4305 [hep-th].
 
\bibitem{Bohr}
  N.~E.~J.~Bjerrum-Bohr, P.~H.~Damgaard, G.~Festuccia, L.~Planté and P.~Vanhove,
  Phys.\ Rev.\ Lett.\  {\bf 121} (2018) no.17,  171601
  doi:10.1103/PhysRevLett.121.171601
  [arXiv:1806.04920 [hep-th]].
\bibitem{Oco}
  D.~A.~Kosower, B.~Maybee and D.~O'Connell,
  JHEP {\bf 1902} (2019) 137
  doi:10.1007/JHEP02(2019)137
  [arXiv:1811.10950 [hep-th]].
\bibitem{Damourperi} 
  T.~Damour and G.~Schaefer,
  Nuovo Cim.\ B {\bf 101}, 127 (1988).
  doi:10.1007/BF02828697
\bibitem{Vanhove1}
  P.~Brax, P.~Valageas and P.~Vanhove,
  Phys.\ Rev.\ D {\bf 97} (2018) no.10,  103508
  doi:10.1103/PhysRevD.97.103508
  [arXiv:1711.03356 [astro-ph.CO]].
  
\bibitem{Vanhove2}
  P.~Brax, P.~Valageas and P.~Vanhove,
  Int.\ J.\ Mod.\ Phys.\ A {\bf 33} (2018) no.34,  1845006.
  doi:10.1142/S0217751X18450069
  
\bibitem{Alvarez}
  L.~Alvarez-Gaume, A.~Kehagias, C.~Kounnas, D.~Lüst and A.~Riotto,
  Fortsch.\ Phys.\  {\bf 64} (2016) no.2-3,  176
  doi:10.1002/prop.201500100
  [arXiv:1505.07657 [hep-th]].

\end{thebibliography}
\end{document}